\documentclass[reprint,amsmath,amssymb,aps,prl]{revtex4-1}

\usepackage{graphicx}
\usepackage{natbib}
\usepackage{xfrac}
\usepackage{epstopdf}

\newcommand{\bra}[1]{\ensuremath{\left<#1\right|}}
\newcommand{\ket}[1]{\ensuremath{\left|#1\right>}}
\newcommand{\unit}[1]{\ensuremath{\,\mathrm{#1}}}
\newcommand{\WV}[1]{\ensuremath{\langle{#1}\rangle_W}}

\begin{document}

\title{Demonstration of Weak Measurement Based on Atomic Spontaneous Emission}

\date{14/7/2013}

\author{Itay Shomroni}
\author{Orel Bechler}
\author{Serge Rosenblum}
\author{Barak Dayan}
\email{barak.dayan@weizmann.ac.il}
\affiliation{Department of Chemical Physics, Weizmann Institute of Science, Rehovot 76100, Israel}

\begin{abstract}
We demonstrate a new type of weak measurement based on the dynamics of spontaneous emission.
The pointer in our scheme is given by the Lorentzian distribution characterizing atomic exponential decay via emission of a single photon.
We thus introduce weak measurement, so far demonstrated nearly exclusively with laser beams and Gaussian statistics, into the quantum regime of single emitters and single quanta, enabling the exploitation of a wide class of sources that are abundant in nature.
We describe a complete analogy between our scheme and weak measurement with conventional Gaussian pointers.
Instead of a shift in the mean of a Gaussian distribution, an imaginary weak value is exhibited in our scheme by a significantly slower-than-natural exponential distribution of emitted photons at the postselected polarization, leading to a large shift in their mean arrival time.
The dynamics of spontaneous emission offer a broader view of the measurement process than is usually considered within the weak measurement formalism.
Our scheme opens the path for the use of atoms and atomlike systems as sensitive probes in weak measurements,
one example being optical magnetometry.
\end{abstract}


\maketitle

The concept of weak measurement was introduced by Aharonov, Albert, and Vaidman ~\cite{Aharonov1988How,Duck1989Sense,[{For recent reviews see }]ShikanoReview,*Dressel2013Understanding} as a part of a time-symmetric formulation of quantum mechanics, by defining a ``weak value'' of an observable relative to both pre- and postselected states of a system.
In a weak measurement, the measured system is only slightly perturbed by the measuring device (pointer), then the system is postselected in a prescribed final state via a regular projective (``strong'') measurement.
The weak value corresponds to a shift of the pointer within the postselected ensemble, analogous to a measured eigenvalue (with no postselection) in a regular quantum measurement.
Aharonov, Albert, and Vaidman showed that the weak value possesses strange characteristics, the most prominent of which is that it may lie outside the eigenvalue spectrum of the measured observable.
In their famous example, a weak measurement of the spin of a spin-\sfrac{1}{2} particle may yield the value of 100~\cite{Aharonov1988How}.

Weak values have been used to address foundational questions in quantum mechanics, such as Hardy's paradox~\cite{Lundeen2009Experimental}, the Leggett-Garg inequality~\cite{Goggin2011Violation,Dressel2011Experimental}, the three-box problem~\cite{Resch2004Experimental}, tunneling time~\cite{Steinberg1995How,*Steinberg1995Conditional}, and for clarifying Heisenberg's uncertainty relation~\cite{Rozema2012Violation}.
They have also been used to map the average trajectories of photons in a two-slit experiment~\cite{Kocsis2011Observing} and to directly measure the quantum wave-function of a photon~\cite{Lundeen2011Direct}.
This prompted a proposal to use weak measurement as an alternative to quantum tomography~\cite{Lundeen2012Procedure}, which has recently been demonstrated~\cite{Salvail2013Full}.

The prospect of a weak value extending beyond the eigenvalue spectrum of an observable is often referred to as amplification, alluding to the possibility of measuring small effects. Recently this approach yielded impressive results, including the observation of the spin Hall effect of light~\cite{Hosten2008Observation}, where a lateral displacement of $\sim 1\unit{\AA}$ of a light beam was amplified by nearly 4 orders of magnitude, permitting observation. Ultrasensitive beam deflection measurement down to $400\unit{frad}$~\cite{Dixon2009Ultrasensitive}, and measurement of Doppler shifts as small as $9\unit{\mu Hz}$ have also been performed~\cite{Viza2013Weak}. Although weak measurement does not improve upon the shot-noise limit, many practical applications are limited by technical noise and weak value amplification may offer substantial improvements in signal-to-noise ratio~\cite{Brunner2010Measuring,Feizpour2011Amplifying}.

To the best of our knowledge, all experimental demonstrations of weak measurements to date employed light as both the measured system and measuring device~\cite{Dixon2009Ultrasensitive,Viza2013Weak,Goggin2011Violation,Dressel2011Experimental,Resch2004Experimental,
Lundeen2009Experimental,Kocsis2011Observing,Salvail2013Full,Hosten2008Observation,
Ritchie1991Realization,Lundeen2011Direct,Wang2006Experimental,Gorodetski2012Weak}.
Applications to condensed-matter systems~\cite{Zilberberg2011Charge,Williams2008Weak,Dressel2012Measuring} and atomic ensembles~\cite{Simon2011Fockstate} have been proposed.
Moreover, the conventional pointer in weak measurement schemes has so far been Gaussian. Indeed the Gaussian mode of a laser beam has been conveniently used for this purpose almost exclusively.
Only a few theoretical studies have considered other possible pointers~\cite{Berry2011Pointer, Puentes2012Weak, Susa2012Optimal, Kobayashi2012Extracting, Hayat2012Enhanced},
including a recent  proposal for probing Fermion interactions by real weak value amplification, which considers a complex Lorentzian pointer~\cite{Hayat2012Enhanced}.
The real weak value amplification properties of real Lorentzian and symmetric exponential pointers have also been theoretically studied~\cite{Berry2011Pointer}.

In this Letter we demonstrate weak measurement of polarization utilizing an atom and its coupling to the environment via spontaneous emission as a pointer.
In contrast to the conventional Gaussian pointers, in our scheme the pointer is a complex Lorentzian, corresponding to the spectral amplitude of spontaneously emitted photons.

We perform an imaginary weak value measurement; i.e., we measure a significant change in the observed exponential decay rate of the atom in the time domain, leading to a large shift in the mean arrival time of the emitted photons.
The amplification in our scheme is exhibited in the fact that a very small energy splitting between atomic levels (smaller than the linewidth) leads to a significant effect at a very short time-scale (doubling of the mean arrival time of the photons).

We now briefly outline the weak measurement formalism.
Standard measurement procedure in quantum mechanics was formulated by von Neumann~\cite{vonNeumann}.
It consists of first preparing the system in a state $\ket\psi$, then coupling it to a pointer, characterized by some initial distribution in $q$ space (i.e., position), localized at $q_0$ with some uncertainty $\delta q$.
The coupling is given by the Hamiltonian
\begin{equation}
H = -g(t)\Delta\,\hat p\,\hat O,
\label{eq:H1}
\end{equation}
where $\hat O$ is the system observable being measured, with eigenvalues $a_i$.
The parameter $\Delta>0$ depends on the experimental apparatus and relates the units of the pointer and observable.
The operator $\hat p$ is the coordinate of the pointer conjugate to $\hat q$.
The normalized function $g(t)$ represents the measurement duration.
It is usually approximated by a $\delta$ function (impulsive interaction) so that all free evolution terms in the Hamiltonian can be neglected during measurement~\cite{Haake1987Overdamped}.

Following the interaction of the system and pointer, given by the evolution operator $e^{-i\int\!H dt}$, the pointer gains a linear phase in its conjugate coordinate $\hat p$, corresponding to a shift in  $q$ space that is proportional to $\hat O$.
The measured value is inferred from the shift of the pointer $\langle\hat q\rangle-q_0$. In a strong measurement the differences between the $a_i$
are much larger than the pointer uncertainty $\delta q$, resulting in well-defined shifts located at $\Delta a_i$.
The value of $\hat O$ is thus determined from a single measurement.

In contrast, in a weak measurement the uncertainty of the pointer is much larger than its shift~\cite{[{Note that this discussion deals with pure states. Mixed states in the context of weak measurements have been discussed, e.g., in }] DiLorenzo2008Weak, *Cho2010Weak}.
The ensuing correlation between the system and pointer is weak, and the system is perturbed only slightly.
The value of $\hat O$ cannot be inferred from the pointer in this case.
Its average $\langle\hat O\rangle$ can still be recovered from many weak measurements on equivalent preparations.
However the true strength of the weak measurement process is evident when postselecting the outcome, by performing a subsequent strong measurement of a different variable of the system and retaining only instances with a given outcome $\ket{\phi}$.
Reference~\onlinecite{Aharonov1988How} introduced the notion of a ``weak value'' of an operator,
$\WV{\hat O} = \bra\phi\hat O\ket\psi/\left<\phi|\psi\right>$ and showed that for real weak values, under some approximations~\cite{Duck1989Sense}, the pointer average following postselection is shifted by $\WV{\hat O}$.
The pointer shift is thus ``amplified'' by postselection when $\left<\phi|\psi\right>\ll 1$.
The amplification is due to constructive interference in the tail of the distribution, and destructive interference elsewhere, tailored through judicious use of postselection~\cite{Duck1989Sense}.
This amplification comes at a cost of reducing the signal due to many instances not passing postselection.

In contrast to strong measurements which yield real eigenvalues, a weak value can be complex.
A complex weak value gives rise to shifts in both canonical coordinates of the pointer~\cite{Jozsa2007Complex},
with a shift in $\hat p$ proportional to $\mathrm{Im}\WV{\hat O} $.
An imaginary weak value is commonly encountered experimentally~\cite{Brunner2010Measuring, Dixon2009Ultrasensitive,Hosten2008Observation,Viza2013Weak}.

\begin{figure}
\includegraphics{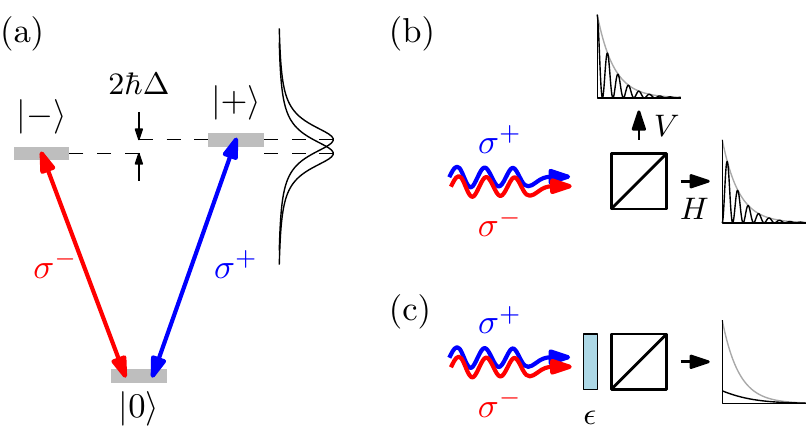}
\caption{(a)~Atomic V system for a weak measurement using atomic line shapes.
The atom is excited from the ground state to a superposition of the upper states by a short pulse at vertical polarization $V=\sigma^+\!+\sigma^-$, and spontaneously decays back to the ground state.
In an external magnetic field, the energies of the upper levels are Zeeman shifted by $\pm\hbar\Delta$.
The two line shapes with width $\Gamma$ are shown, overlapping when $\Delta\ll\Gamma$.
(b)~For $\Delta\gg\Gamma$, interfering the two circular polarizations with a PBS leads to time-domain quantum beats.
(c)~A weak measurement corresponds to the case in which $\Delta\ll\Gamma$ and a linear polarization that is rotated by $\epsilon$ from $H$ is postselected, with $\Delta/\Gamma\ll\epsilon\ll1$. The result is an exponentially decaying signal with a longer-than-natural decay time.}
\label{fig:1}
\end{figure}

We now turn to spontaneous emission.
For an atom at the excited state, the spectral amplitude distribution of a spontaneously emitted photon after a long time is~\cite{[{See, e.g., }][{, p. 195}] CohenTannoudji}
$L(\omega) \propto 1/[1+2i(\omega-\omega_0)/\Gamma]$, with $\omega_0$ the resonant frequency of the transition and $1/\Gamma$ the excited state lifetime.
In the time domain this corresponds to the probability of emitting a photon $\propto \Theta(t)e^{-\Gamma t}$ with  $\Theta(t)$ the Heaviside step function.
The spectral intensity of the emitted light is given by $|L(\omega)|^2$, a Lorentzian with full width at half maximum (FWHM) $\Gamma$.

Our weak measurement scheme uses an atom with a V-level structure as depicted in Fig.~\ref{fig:1}(a). The two upper levels $\left|\pm\right>$, with magnetic quantum number $m=\pm1$, are excited from the $m=0$ ground state by opposite circular polarizations $\sigma^{\pm}$.
An external magnetic field shifts the upper levels by $\pm\hbar\Delta$ due to the Zeeman effect [Fig.~\ref{fig:1}(a)].
The atom is excited instantaneously at $t=0$ with a linearly polarized pulse into a superposition $\ket\psi = \frac{1}{\sqrt{2}}\left(\ket+ +\ket-\right)$, and decays spontaneously back to the ground state. 

For simplicity, henceforth we consider photons emitted in the direction of the magnetic field, for which the one-to-one mapping between their (circular) polarization and the excited state of the atom allows us to treat the photon polarization and the atomic excited state as the same system.
Coherence in this system requires the absence of any which-path information in the final states of both the atom and the photon.
While the V-system configuration ensures this is fulfilled by the atomic state, which-path information must be erased from the polarization state of the emitted photon.
This can be done by using a polarizing beam splitter (PBS) to project its state on a basis of linear polarizations.

When $\Delta\gg\Gamma$, such a system gives rise to quantum beats~\cite{Haroche1976Beats,*Haroche1973Hyperfine, *[{Also note the existence of the Hanle effect in such a system, see e.g., }] Budker2002Resonant}, in which the photon detection events at the two output ports of the PBS are proportional to 
$e^{-\Gamma t}\cos^2(\!\Delta\,t)$ and $e^{-\Gamma t}\sin^2(\!\Delta\,t)$  [Fig.~\ref{fig:1}(b)].

In contrast, a weak measurement corresponds to the entirely different regime where $\Delta\ll\Gamma$, i.e., a very weak magnetic field.
The beat frequency is then obscured by the much faster exponential decay [Fig.~\ref{fig:1}(c)].
In the frequency domain this corresponds to two Lorentzians displaced from one another by much less than their widths, in accordance with the weak measurement picture, with the natural line shape serving as a pointer measuring polarization.

Note that this is not a standard von Neumann measurement of the type of Eq.~\eqref{eq:H1}.
The atom-field interaction is not impulsive, and we cannot neglect the atom and field free evolution terms in the total Hamiltonian.
Indeed it is the combination of these terms together with the interaction term that determine the spectral and time behavior of the system.
Correspondingly there is no initial pointer that undergoes a shift.
Nevertheless the same interference effect that characterizes weak measurement holds.
We can treat this formally by defining an effective evolution operator for $t\rightarrow\infty$ by $U(\omega) = e^{i\Delta\hat\sigma_z\hat\tau}L(\omega)$ where $\hat\sigma_z = \ket+\bra+ - \ket-\bra-$ corresponds to the measured observable in Eq.~\eqref{eq:H1}.
Mathematically, we use $\hat\tau=-i\,d/d\omega$ to generate the shift in the spectrum. Note that $\hat\tau$ is not intended to represent a quantum mechanical operator.

The emitted photon is postselected for linear polarization at an angle of $90^{\circ}+\epsilon$ to the prepared state, with $0<\epsilon\ll 1$, namely at $\ket\phi = \frac{1}{\sqrt{2}}\left(e^{-i\epsilon}\ket+ - e^{i\epsilon}\ket-\right)$.
The weak value $\WV{\hat\sigma_z}\equiv\bra\phi\hat\sigma_z\ket\psi/\left<\phi|\psi\right> = i\cot\epsilon$ is purely imaginary, which leads us to look for the effect on the pointer in the time domain.
Following postselection, the state of the pointer is $\bra\phi U(\omega)\ket\psi$.
In the time domain the resulting photon probability distribution is
\begin{equation}
P(t) = C e^{-\Gamma t}\sin^2(\Delta\,t+\epsilon),
\label{eq:P}
\end{equation}
with $C$ a normalization constant.
Amplification in weak measurement requires
$\Delta/\Gamma \ll\epsilon\ll 1$~\cite{Duck1989Sense}.
In this regime
\begin{equation}
P(t) \simeq C \epsilon^2 \exp\left[-\left(1-\frac{2\Delta}{\epsilon\Gamma}\right)\Gamma t\right].
%P(t) \simeq C \epsilon^2 \exp\left[-\left(1-2\Delta/\epsilon\Gamma\right)\Gamma t\right].
\end{equation}
The result of postselection is thus alteration of the decay constant by the factor in parentheses (at the cost of lowering the signal by $\epsilon^2$), leading to a shift in the mean of the normalized pointer in the time domain of
\begin{equation}
\int\!\! t P(t)dt - \int\!\!t P_0(t)dt \simeq \frac{2\Delta\cot\epsilon}{\Gamma^2} =
2\Delta \mathrm{Var} P_0 \mathrm{Im}\WV{\hat\sigma_z}
\label{eq:shift}
\end{equation}
with $P_0(t)\equiv\Gamma e^{-\Gamma t}$ the ``initial'' pointer (with a mean of $1/\Gamma$).
This result agrees with the analysis performed by Jozsa (Eq.~10 of Ref.~\onlinecite{Jozsa2007Complex}), and in particular is equivalent to the shift in the mean of Gaussian pointers~\cite{[{We note that other theoretical works treated the concept of arrival time in the context of weak measurement, although not in a context that relates to our work: }] Ruseckas2002Weak, *Ahnert2004Weak}.

\begin{figure}
\includegraphics{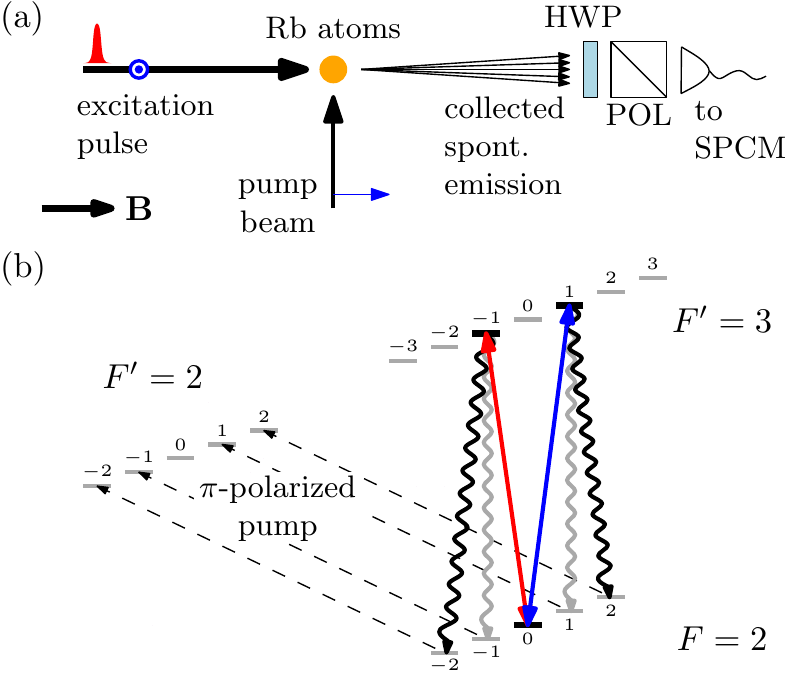}
\caption{(a)~Experimental setup. Ultracold $^{87}$Rb atoms are released from a MOT and a small magnetic field $\mathbf{B}$ is applied. The atoms are pumped to the $\ket{F=2\ m=0}$ state using a $\pi$-polarized pump beam (for which this state is a dark state due to selection rules). The atoms are then excited by a $4.2\unit{ns}$ pulse linearly polarized perpendicular to the magnetic field, inducing $\sigma^\pm$ transitions only. Spontaneous emission in the direction of the magnetic field is collected into a multimode fiber.
Time is measured from the triggering of the excitation pulse.
%Blue arrows denote linear polarization direction.
HWP, half-wave plate; POL, Glan-Thompson polarizer; SPCM, single-photon counting module.
(b)~$^{87}$Rb levels used in the experiment. The V system transitions are shown in straight solid arrows. Gray wavy arrows are $\pi$-decay paths eliminated by our observation direction. Black wavy arrows are undesired $\sigma$ decays, which add incoherent background to our measurements, at a level of ${}\sim 12\%$ of the total signal~\cite{SuppInfo}.
The $\pi$-polarized optical pumping is shown as dashed lines.}
\label{fig:experimental}
\end{figure}

This analysis highlights the fact that a weak measurement need not rely on the {\it a priori} existence of a pointer, as long as we can define a shift proportional to the measured parameter $\Delta$.
Alternatively, we can rigorously treat the atom as a \emph{quantum} pointer, acquiring macroscopic properties through its coupling to the environment (via spontaneous emission)~\cite{Haake1987Overdamped}.

%%% EXPERIMENTAL

As a proof-of-principle, we demonstrate this effect with an ensemble of ultracold $^{87}$Rb atoms ($1/\Gamma=26\unit{ns}$) released from a magneto-optical trap (MOT).
The experimental setup is depicted in Fig.~\ref{fig:experimental}(a).
At the start of the experiment the atoms are cooled by polarization-gradient cooling to $\sim\! 10\unit{\mu K}$, the trap is turned off and a small bias magnetic field is applied.

As shown in Fig.~\ref{fig:experimental}(b), $^{87}$Rb is not the ideal atomic species for this experiment due to its multitude of energy levels.
Therefore, we created an effective V system among these levels by choosing the subset $\ket 0\equiv\ket{F=2\ m=0}$ and $\ket\pm\equiv\ket{F'=3\ m=\pm 1}$, pumping the atoms to the $m=0$ ground state before each measurement (Fig.~\ref{fig:experimental}). The atoms were then excited from $\ket 0$ to $\ket{+}+\ket{-}$ by a $4.2\unit{ns}$ pulse, linearly polarized perpendicular to the magnetic field.

We made sure our measurements were dominated by the signal from this effective V system by tailoring the exciting-pulse duration to minimize excitation to states outside this V system, and collecting photons emitted only at a small angle from the magnetic field direction to minimize the contribution of undesired decay channels (see Fig.~\ref{fig:experimental}(b) and Supplemental Material for additional details~\cite{SuppInfo}). The remaining incoherent background (measured to be at a level of ${}\sim\! 12\%$ of the coherent signal) was estimated by reference measurements at $\epsilon=0$ that were interleaved with the data measurements, and subtracted from our results~\cite{SuppInfo}.

The spontaneously emitted photons were coupled through a multimode fiber into a single-photon counting module (SPCM).
A combination of a half-wave plate and a polarizer in front of the fiber realized the postselection (Fig.~\ref{fig:experimental}).
The SPCM events were collected by a multiple-event time digitizer (FAST-ComTec MCS6). The time digitizer was triggered together with the excitation pulse, giving a time base with a resolution of $100\unit{ps}$~\cite{SuppInfo}.

%%% RESULTS

\begin{figure}
\includegraphics{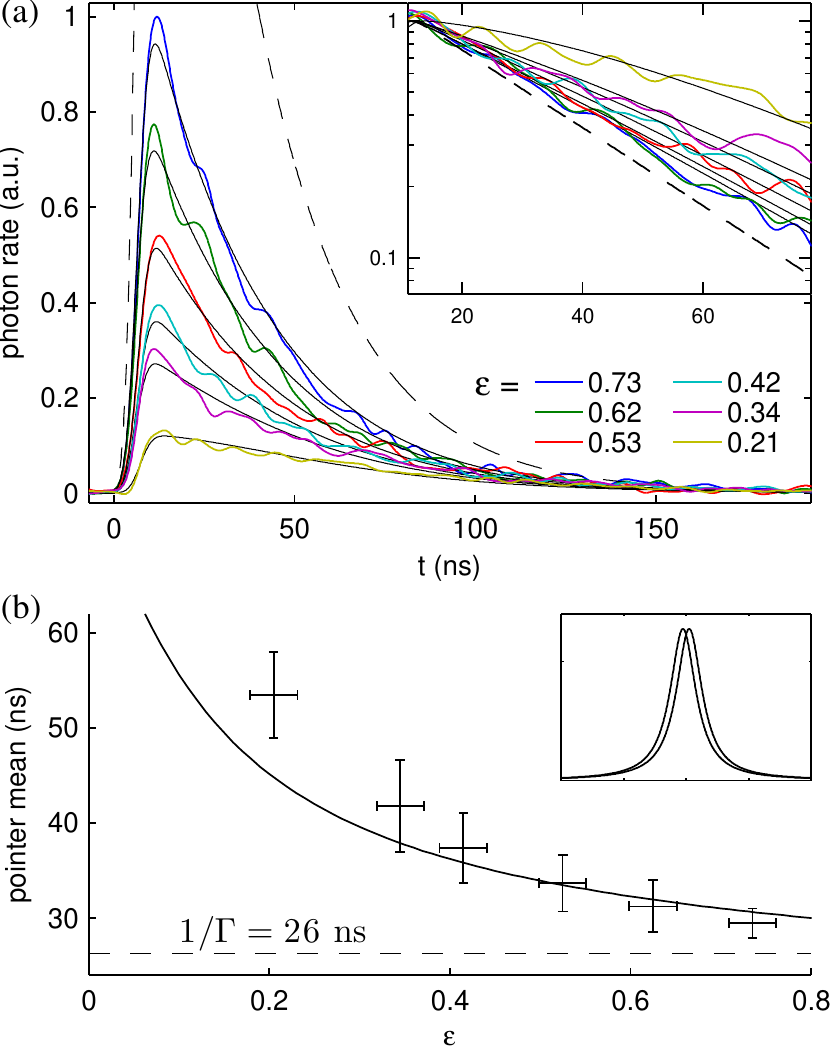}
\caption{Experimental results. (a)~Temporal distribution of arrived photons for various degrees of postselection $\epsilon$.
The case of no postselection (natural decay with lifetime $1/\Gamma$) is shown in dashed black line, for comparison.
The solid black lines are fits of the data to Eq.~\eqref{eq:P} with no free parameters except a scale factor.
The shift of the mean to later times with smaller values of $\epsilon$ is evident.
Data were low-pass filtered using a Gaussian of $4.2\unit{ns}$ full width.
Inset: logarithmic presentation of the decaying part of the signal.
(b)~Mean arrival time computed from the experimental data for various values of $\epsilon$.
The solid curve is the theoretical mean based on exact solution for $\Delta=600\unit{kHz}$.
The dashed horizontal line indicates the mean that corresponds to the natural lifetime.
Inset: The two corresponding line shapes with separation equal to \sfrac{1}{5} FWHM.}
\label{fig:results}
\end{figure}

The results are shown in Fig.~\ref{fig:results}.
We have used $\Delta=600\unit{kHz}$ to illustrate a weak measurement with $6\unit{MHz}$ FWHM Lorentzians, as shown in the inset of Fig.~\ref{fig:results}(b).
The considerable overlap of the states ensures that we are in the weak measurement regime.
Figure~\ref{fig:results}(a) shows the distribution of photon arrival times (the pointer) for various postselection parameters $\epsilon$, after the incoherent background has been subtracted.
A fit of the data to Eq.~\eqref{eq:P} convolved with the excitation pulse shape, with no free parameters except a scale factor, shows excellent agreement, as shown by the black lines.
Figure~\ref{fig:results}(b) shows the mean photon arrival times computed from our data vs~$\epsilon$,
compared to the theoretical value based on the mean of Eq.~\eqref{eq:P}
(accounting for the excitation pulse shape).
There seems to be a small systematic shift between the measured values and the theoretical curve, that may be attributed to the effect of imperfect subtraction of the (slightly varying) incoherent background, especially in small values of $\epsilon$.

%%% CONCLUSION

To conclude, by introducing a new type of complex-Lorentzian pointers with the exponential decay dynamics that characterize coupling to the environment, we apply the principles of weak measurement to a new class of light sources that are abundant in nature, like spontaneous emission from atoms and atomlike systems, and even NMR~\footnote{I.~Shomroni~\emph{et al.} (to be published). We note that in NMR the quadratures of the magnetic field replace the polarization degree of freedom.}.
Our scheme, which is applicable both to ensembles and single emitters, is inherently suitable for true atomic V systems with narrow linewidths, e.g. the $^1S_0\rightarrow{}^3P_1$ optical transition in alkaline earth atoms, such as $^{88}$Sr, $^{40}$Ca, and $^{24}$Mg, offering a single-step, all-optical measurement of energy-level shifts with sensitivity potentially better than $1\unit{Hz}/\!\unit{\sqrt{Hz}}$.

%%% ACKNOWLEDGEMENTS

\begin{acknowledgments}
We wish to thank L.~Vaidman for fruitful discussions.
Support from the Israeli Science Foundation and the Joseph and Celia Reskin Career Development Chair in Physics  is acknowledged.
This research was made possible in part by the historic generosity of the Harold Perlman Family.
\end{acknowledgments}

\bibliography{weak}

%%% SUPPLEMENTAL MATERIAL

\clearpage

\onecolumngrid

\section{Supplemental Material}

\emph{Optical Pumping.}
As mentioned in the main text, prior to every excitation we pumped the atoms to the $\ket 0\equiv\ket{F=2\ m=0}$ state.
This pumping was accomplished by a pump beam tuned to the $F=2\rightarrow F'=2$ transition and polarized along the magnetic field, resulting in $\pi$-transitions only (Fig.~2 of the main text).
The state $\ket 0$ is a `dark state' for the pump beam due to selection rules. All other $F=2$ states are excited and decay either to $\ket 0$ or to the $F=1$ ground state.\\

\emph{Incoherent background due multiple excitation and decay paths.}
Following the pumping, the atoms were excited from $\ket 0$ to $\ket{+}+\ket{-}$ by a $4.2\unit{ns}$ pulse, linearly polarized perpendicular to the magnetic field. The atoms can subsequently decay to any sublevel of the $F=2$ manifold, whereas only decays to $\ket 0$ are part of the desired V-system and contribute to the coherent signal  (Fig.~2b of the main text).
The contribution of undesired decay channels can be minimized by collecting only photons emitted to a small angle around the direction of the magnetic field, in which only $\sigma$ photons are emitted, thus excluding photons emitted by $\pi$-transitions to $\ket{F=2\ m=\pm 1}$.
However, $\sigma$ photons from undesired decay to $\ket{F=2\ m=\pm 2}$ are still detected.

The reason we chose this particular configuration to approximate an ideal V-system, is the fact that the dipole matrix elements for decay into $m_F=0$ are $\sqrt 6$ times larger than those for $m_F=\pm 2$.
Moreover, in the basis of linear polarizations (in which there is no which-path information about the origin of the emitted photon) the amplitudes of the two decay channels into $m_F=0$ add coherently, since they end up in the same final state.
This makes a photon-detection event from this process 12 times more likely than from decays to $m_F\pm 2$, which end up at different ground states, and accordingly add incoherently. This configuration leads to the presence of an incoherent background at a level of $\sim 8\%$ of the total signal. 

Other mechanisms may further increase this incoherent background. First, our spectrally-wide pulse could excite the atom also to the $F'=2$ manifold, $267\unit{MHz}$ away from the $F'=3$ manifold.
To minimize this effect we chose the duration of our approximately square pulse to be $4.2\unit{ns}$, leading to a sinc-shaped spectrum which ideally drops to zero at $267\unit{MHz}$ from the center frequency. An additional source for such background are room-temperature atoms passing through the excitation region, which are not resonant with the pump beam (due to Doppler shift), but are nonetheless excited by the spectrally-wide excitation pulse. Overall, our measured incoherent background was at a level of $\sim 12\%$ of the total signal. Note that this incoherent background would not exist in a true V-system.

We also note that due to geometrical limitations, in our setup we detected photons emitted at a slight offset of $\sim4.5^\circ$ from the direction of the magnetic field. The effect of this small offset on our results was both calculated and measured to be completely negligible.\\

\emph{Compensating for SPCM after-pulsing and dead time.}
Our SPCM was measured to have after-pulsing probability of $\sim 2\%$ following each detection event (at the end of the detector's dead time, $\sim 52\unit{ns}$). This effect is especially detrimental to our measurement, since it effectively increases the observed lifetime by adding $\sim 2\%$ of the peak signal (including background) at roughly twice the natural lifetime, where the actual signal is $\sim 1/e^2$ weaker.

To prevent this, the raw-data photon detection events were scanned to remove any event that occurred less than $115\unit{ns}$ after its predecessor, thereby eliminating both orders of after-pulsing. This process was then taken into account by assuming a dead time of $115\unit{ns}$ instead of $52\unit{ns}$. 

To compensate for the effect of dead time, the remaining detection events were then collected to histogram with $100\unit{ps}$ time bins, and a correction factor was calculated for each bin, according to the average probability that the detector is in a dead time due to a previous photon detection event in the last $115\unit{ns}$.\\

\emph{Multiple Scattering.}
Multiple scattering in the atomic cloud may prevent the detected photon from being the first one emitted after state preparation, thus adding to the incoherent signal. In our experiment this effect was practically eliminated by reducing the density of the atomic cloud during the measurement, keeping the optical density at a level of a few percent. This was verified by measuring a natural lifetime that agreed with the theoretical value of $1/\Gamma$ up to $\pm4\%$. Analytic and numeric calculations confirmed that the effect of such a low level of multiple scattering on our results is completely negligible.\\

\begin{figure}
\includegraphics[scale=0.8]{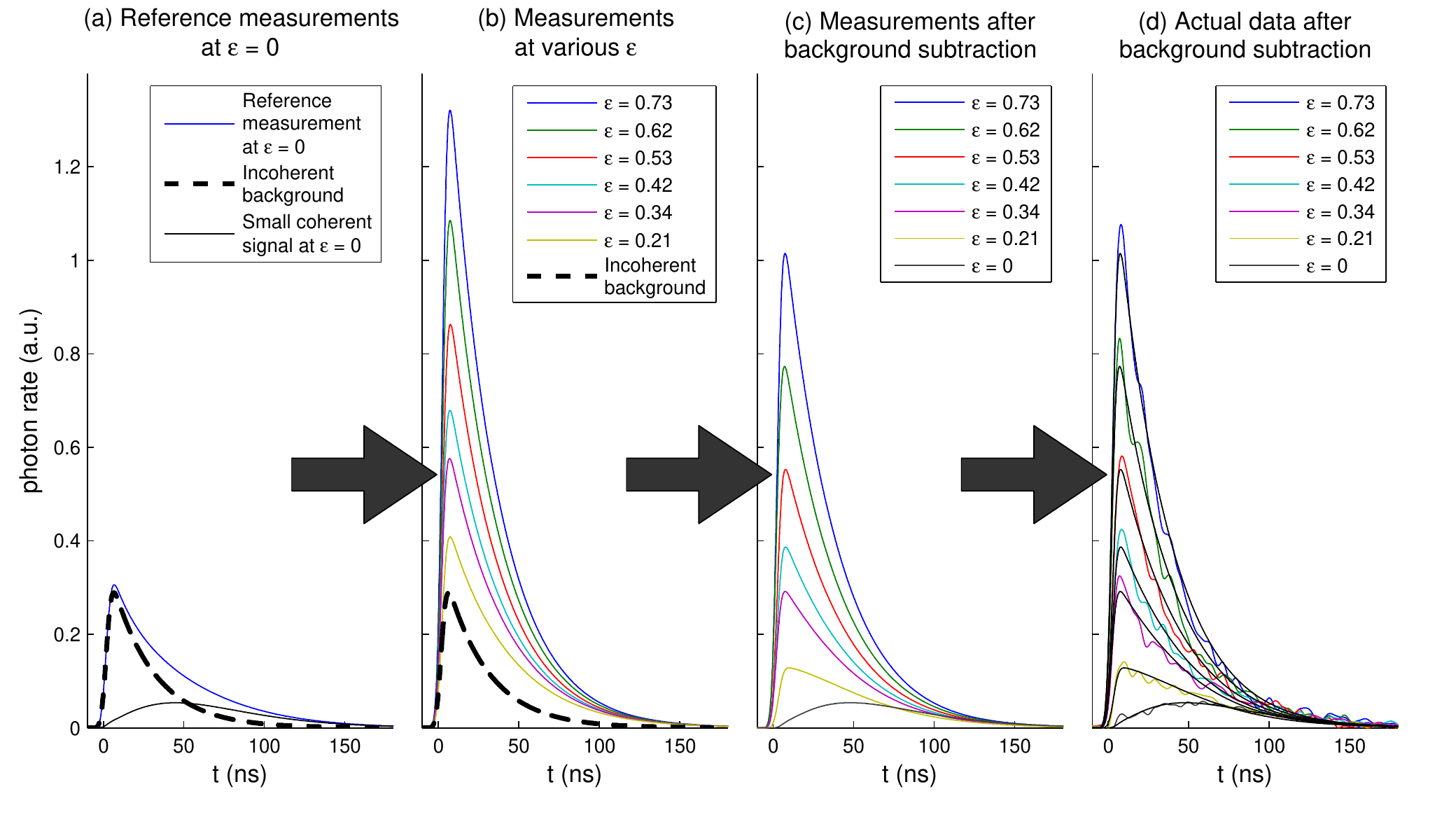}
\caption{Measurement and data analysis procedure (see text).}
\label{fig:supplemental}
\end{figure}

\emph{Reference Measurements at $\epsilon=0$.}
Our entire measurement and analysis procedure is shown in Fig.~\ref{fig:supplemental}. As described above, the fact that our atomic configuration was not a perfect V-system resulted in the presence of an incoherent background in all our measurements. To remove this background we performed reference measurements at $\epsilon=0$, at which the incoherent background is most prominent. Ideally, the reference measurements at $\epsilon=0$ should have been performed with $B=0$, at which the coherent signal is zero throughout the entire measurement, leaving only the incoherent background. However, changing the magnetic field affects many parameters in our system, the pumping efficiency in particular, thereby preventing using this measurement as an accurate reference. Instead we used a fit to Eq.~(2) to isolate the component of pure exponential decay, removing the additional small coherent signal (Fig.~\ref{fig:supplemental}a). The fact that the coherent and incoherent signals are drastically different (with the coherent signal starting at zero, where the incoherent signal is maximal) makes this mathematical procedure accurate and reliable.

The reference measurements were interleaved with the other measurements at $\epsilon\neq 0$. This allowed the exponential component to be isolated from two reference measurements for every $\epsilon\neq 0$ measurement---the one before and the one after. This exponential component was then subtracted from the $\epsilon\neq 0$ data, as shown in Fig.~\ref{fig:supplemental}b--c. Beyond improving the SNR and accuracy of the background measurement, the consistency of the two reference measurements was used to verify the stability of our system during the $\epsilon\neq 0$ measurement. Fig.~\ref{fig:supplemental}d shows the experimentally measured signals for all values of $\epsilon$, including 0.

\end{document}